\documentclass[12pt]{article}
\usepackage{amssymb}
\usepackage{psfig}
\newcommand{\beq}{\begin{equation}}
\newcommand{\eeq}{\end{equation}}
\newcommand{\beqn}{\begin{eqnarray}}
\newcommand{\eeqn}{\end{eqnarray}}
\newcommand{\bea}[1]{\beq\begin{array}{#1}}
\newcommand{\eea}{\end{array}\eeq}
\newcommand{\eq}[1]{(\ref{#1})}

\newcommand{\dual}[1]{{}^{*}{#1}}

\newcommand{\tr}{\mathop{\rm Tr}}

\newcommand{\diff}{\partial}

\newcommand{\cD}{{\cal D}}

\newcommand{\AP}[3]{{\it Ann. Phys. }{\bf #1} (#2) #3}
\newcommand{\NP}[3]{{\it Nucl. Phys. }{\bf #1} (#2) #3}
\newcommand{\NPPS}[3]{{\it Nucl. Phys. Proc. Suppl. }{\bf #1} (#2) #3}
\newcommand{\PL}[3]{{\it Phys. Lett. }{\bf #1} (#2) #3}
\newcommand{\PRL}[3]{{\it Phys. Rev. Lett. }{\bf #1} (#2) #3}
\newcommand{\PRep}[3]{{\it Phys. Rep. }{\bf #1} (#2) #3}
\newcommand{\PR}[3]{{\it Phys. Rev. }{\bf #1} (#2) #3}

\newcommand{\PTP}[3]{{\it Prog. Theor. Phys. }{\bf #1} (#2) #3}
\newcommand{\PTPS}[3]{{\it Prog. Theor. Phys. Suppl. }{\bf #1} (#2) #3}

\begin{document}
\date{}
\title{Towards Abelian-like formulation of\\
the dual gluodynamics.
\vskip-40mm
\rightline{\small ITEP-TH-28/00}
\vskip 40mm
}
\author{M.N. Chernodub$^{\rm a}$, F.V.~Gubarev$^{\rm a,b}$,
M.I. Polikarpov$^{\rm a}$,
V.I.~Zakharov$^{\rm b}$ \\
\\
$^{\rm a}$ {\small\it Institute of Theoretical and  Experimental Physics, B.Cheremushkinskaya 25, Moscow,}\\
           {\small\it  117259, Russia}\\
$^{\rm b}$ {\small\it Max-Planck Institut f\"ur Physik, F\"ohringer Ring 6, 80805 M\"unchen, Germany}
}
\maketitle
\thispagestyle{empty}
\setcounter{page}{0}
\begin{abstract}\noindent
We consider gluodynamics in case when both color and magnetic charges are present.
We discuss first short distance physics, where only the fundamental $|Q_m|=1$ monopoles
introduced via the `t~Hooft loop can be considered consistently. We show that at short
distances the external monopoles  interact as pure Abelian objects. This result can be
reproduced by a Zwanziger-type Lagrangian with an Abelian dual gluon. We introduce also
an effective dual gluodynamics which might be a valid approximation at distances
where the monopoles $|Q_m|=2$ can be considered as point-like as well. Assuming the monopole
condensation we arrive at a model which is reminiscent in some respect of the Abelian Higgs
model but, unlike the latter leaves space for the Casimir scaling.
\end{abstract}

\newpage

\section{Introduction.}

Constructing gauge theories where the same gauge bosons interact with
both magnetic end electric charges is a long standing problem. In case of electrodynamics,
a particular  solution is provided by the Zwanziger formalism \cite{zwanziger}. In this formalism,
one introduces formally two
different gauge fields, $A_{\mu}, B_{\mu}$ interacting with the electric and magnetic charges, respectively.
To avoid the doubling of the degrees of freedom one imposes, however, the constraint that the field strength
tensors build up on the potentials $A_{\mu}$  and $B_{\mu}$ are dual to each other:
\beq
F_{\mu\nu}(A)~=~\dual F_{\mu\nu}(B) ~=~
\frac{1}{2}\varepsilon_{\mu\nu\lambda\rho} F_{\lambda\rho}(B)\,.
\eeq
If one resolves the constraint and works in terms of a single potential, $A_{\mu}$, then the string
singularities naturally emerge.

A direct evaluation of the radiative corrections within the Zwanziger formalism brings in a paradoxical
conclusion that the Dirac quantization condition for the product of the electric and magnetic coupling,
$e\cdot g= 2\pi$, gets violated (for a review and further references see \cite{blagoevic}). The origin
of this mismatch can be traced back to violation, within the perturbative approach of the Dirac
veto which forbids direct interaction of particles with the Dirac strings. This turns out to be not true
for virtual particles. There exist approaches, non-perturbative in nature, which respect the Dirac veto and
then one can see, indeed, that the Dirac quantization condition survives the radiative corrections, for
reviews see \cite{blagoevic,coleman}. At this point, a more detailed analysis of the Dirac strings is
required so that one leaves, in a way, the ground of pure field theories.

Our interest here is in constructing gluodynamics which would incorporate interaction both with color and
magnetic charges. The history of this topic is also rich, see, e.g.,
\cite{coleman,baker,brambilla,ichie,kondo} and references therein. It is worth emphasizing from the very
beginning that there are two different types of monopoles which can be considered in case of gluodynamics.
Monopoles with the minimal magnetic charge, $|Q_m|=1$ can be introduced via the 't~Hooft loop
\cite{tHooft-loop}. In the classical limit they correspond to classical solution with infinite action which
are nothing else but the Dirac monopoles. These monopoles can be considered only as external sources
to probe the vacuum state of the gluodynamics. The monopoles with $|Q_m|=2$,
on the other hand, do not exist as stable classical
solutions \cite{Brandt-Neri}. Nevertheless just these monopoles were observed and studied in great detail
in the lattice simulations, for review and further references see, e.g., \cite{map}. A crucial step to
define $|Q_m|=2$ monopoles is the partial gauge fixing, which leaves a remaining  $U(1)$ invariance
\cite{u1}. Then the monopoles are defined with respect to this $U(1)$.

The studies of $|Q_m|=2$ monopoles strongly support the dual-superconductor mechanism of confinement
\cite{confinement} for the heavy quarks in the fundamental representation. Namely, the quark potential
grows linearly at large distances:
\beq
\lim\limits_{r\to\infty} V_{q\bar{q}}(r) ~=~ \sigma r\,,
\eeq
and the monopole contribution to $V_{q\bar{q}}(r)$ almost saturates the total string tension
$\sigma$ \cite{map}. The mechanism assumes condensation of the monopoles and, what is most important
for us, is borrowed from the electrodynamics of superconductors, i.e. is Abelian in nature. As a result,
the dual formulation of the gluodynamics (see, e.g., \cite{coleman,baker,brambilla,ichie,kondo})
effectively reduces to the same $U(1)$ formalism. The procedure is in two steps. First, one assumes
Abelian dominance in the infrared limit of the gluodynamics itself. Then one constructs the effective $U(1)$
theory a la Zwanziger keeping only these Abelian degrees of freedom. Thus, the crucial step is the
Abelian dominance. There are various ways to introduce this dominance in the field theoretical
language. In particular, the approach of Ref. \cite{baker} is to start with an effective theory for
the dual gluons and to assume that it is similar to the Georgi-Glashow model. Thus, one postulates
the existence of dual gluons, Higgs fields in the adjoint representation and the Higgs mechanism
which reduces the symmetry of the dual world\footnote{
For simplicity we will consider $SU(2)$ case throughout the paper.
Generalizations to $SU(N)$ are straightforward.
} from $SU(2)$ to $U(1)$. As a result, only the corresponding $U(1)$ gauge field interacts with the quarks in
the infrared. In other approaches \cite{brambilla,ichie,kondo} one assumes that in a particular
gauge the masses of the ordinary gluons charged with respect to the remaining $U(1)$ are larger than
the mass of the "neutral" gluon. This "neutral" particle interacts both with color and magnetic charges
as in the Ginzburg-Landau theory.

The dual Ginzburg-Landau theories based on the Abelian dominance in gluodynamics are successful
numerically as far as interaction of the heavy quarks in the fundamental representation is concerned.
However, as was emphasized in Ref.~\cite{greensite}, the data on interaction of the quarks in higher
representations pose a very serious challenge to this picture.
The point is that at intermediate distances (which cover in fact almost all the distances measured so far)
the potential between higher representation sources consists of linear and Coulomb-like pieces.
Moreover, the coefficients in front of the linear and Coulomb terms are proportional to each
other \cite{bali}:
\beq
\label{casimir}
V_T(R)~\approx~ - T(T+1)\;{\alpha_s\over \pi~R}~+~T(T+1)\;\sigma\,R \,,
\eeq
where $T$ labels the representation of the sources and $\sigma$ is independent of $T$. Then the crucial
observation is that in an Abelian-dominant vacuum the quarks which are neutral with respect to the $U(1)$ in
question escape any interaction and there can be no linear potential like (\ref{casimir}). This is true
independent of the form of the interaction in the Abelian sector so that the Casimir scaling
(\ref{casimir}) seems fatal for the dual Ginzburg-Landau approach to the gluodynamics.
More generally, Eq.~(\ref{casimir}) provides a crucial test for any model of confinement
\cite{Simonov}.

In this paper, we revisit the problem of constructing the dual gluodynamics by unifying the
consideration of magnetic monopoles with $|Q_m|=1$ and $|Q_m|=2$. The point is that the fundamental
monopoles with $|Q_m|=1$ are introduced via the 't~Hooft loop and are point-like in the continuum
limit. To describe their interaction we need, therefore, understanding of the dual gluons on the
fundamental level. The basic observation is that classically the fundamental $|Q_m|=1$ monopoles correspond
to $U(1)$ solutions \cite{U1-classification}. Therefore we introduce a dual gluon which is
Abelian on the fundamental level. There is no assumption on the Abelian
dominance whatsoever at short distances and the Abelian nature of the dual gluon
is imposed on us by the well-known fact \cite{coleman,U1-classification} that the monopole
solutions of the Yang-Mills theory are the same $U(1)$ solutions whose embedding into the $SU(2)$
is arbitrary. We derive a Zwanziger-type Lagrangian which unifies standard gluons interacting
with the color and a $U(1)$ dual gluon interacting with external magnetic charges. Then it is a kind of
a gauge freedom how to embed the magnetic $U(1)$ into $SU(2)$. There is no violation of the
color since one averages over all possible embeddings or restricts oneself to evaluation of
gauge-invariant quantities, i.e. the Wilson loops.

We check the validity of this approach by evaluating the running of the coupling
governing the Coulomb-like interaction of the external monopole-antimonopole pair.
As expected, it does run as $1/g^2(r)$ , i.e. preserving the Dirac quantization
condition. However this is true only if the ultraviolet cut-off is introduced
in a way consistent with the Dirac veto. We first postulate such rules and then
check that they do follow directly from an analysis of the continuum analog
of the 't~Hooft loop operator worked out in \cite{main}.

Turning then to physics of larger distances, or physics of confinement we add to the Zwanziger
Lagrangian a phenomenological piece ensuring Higgs mechanism in the dual sector. At this stage, one
can assume that Abrikosov-like vortices with external quarks at the end points are formed. These
solutions are pure Abelian of course but there is no violation of the color symmetry, the same as
at the short distances. Whether the Casimir scaling (\ref{casimir}) is observed or not becomes now
a dynamical question since the string tension, $\sigma_T$ is a function of the phenomenological
parameters, that is of the masses of the dual gluon $m_V$ and of the Higgs particle, $m_H$:
$$
V_T(R)~\approx~ - T(T+1)\;{\alpha_s\over \pi~R}~+~T(T+1)\;\sigma_T(m_H,m_V)\,R \,.
$$
We argue that the Casimir scaling holds in the London limit, $m_H\gg m_V$.
Thus, the approach developed does allow to reconcile the Ginzburg-Landau type
mechanism of confinement with the Casimir scaling.

On the theoretical side, however, there is an important  limitation on the
use of the effective Lagrangian. Namely, upon accounting for the monopole
condensation one cannot treat consistently quantum effects. The reason is
that to impose the Dirac veto for virtual particles becomes now an
intractable problem. Indeed, it was not simple already in case of radiative
corrections to external monopole interactions (see above). And if the
monopoles are condensed the Dirac strings are "everywhere".

The paper is organized as follows. In Section~\ref{Classification} we briefly review the classification
of monopoles in non-Abelian gauge models \cite{coleman,U1-classification,Z2-classification}.
Following the construction of Ref.~\cite{main} in Section~\ref{cont-monopoles} we introduce the notion of
$|Q_m|=1,2$ monopoles in the continuum $SU(2)$ gluodynamics. Then in Section~\ref{Z_2-short-distances}
the fundamental $|Q_m|=1$ monopoles are considered. In particular, we obtain the interaction
potential of the monopole-antimonopole pair and check the running of the coupling.
In Sections~\ref{Z_2-1}, \ref{Z_2-2} we derive the Zwanziger-type Lagrangian for $SU(2)$ gluodynamics
and reconsider the $|Q_m|=1$ monopole interaction in the new framework. Section~\ref{Effective-Model}
is devoted to discussion of $|Q_m|=2$ monopole dynamics at large distances. We argue that within
our approach the assumption of monopole condensation allows to reconcile the dual superconductor
confinement mechanism with the Casimir scaling effect. Our conclusions are summarized in
Section~\ref{Conc}.

\section{Classification of the monopoles}
\label{Classification}

Since we are going to describe interaction of non-Abelian monopoles, the natural
starting point is their classification.
In the literature one can find a few classifications
which look different. In this section we review two particular approaches, namely
the dynamical classification of Ref.~\cite{U1-classification} and topological one
formulated in Ref.~\cite{Z2-classification} (for review see \cite{coleman}).
In the former case the monopoles are classified with respect to $U(1)$ subgroups,
while the latter approach utilizes the global structure of the gauge group, namely
the non-triviality of the gauge group center.
Therefore, we refer to them as $U(1)$ and $Z_2$ classifications respectively.
We conclude that the $Z_2$ classification  applies in fact at short distances and,
in our case, is relevant to the monopoles introduced via the 't~Hooft loop. On the other
hand, the $U(1)$ classification is relevant to monopoles
existing in the vacuum state of the gluodynamics.

The $Z_2$ classification \cite{Z2-classification} is based entirely on topological arguments,
according to which independent types of monopoles are in one-to-one
correspondence with the first homotopy group of the gauge group. Therefore, there are
apparently no magnetic monopoles at all in the $SU(2)$ case. Nevertheless, for
the gauge theory without matter fields in the fundamental representation (quarks)
the actual gauge group is $SO(3)=SU(2)/Z_2$ and there exists a single
topologically non-trivial monopole solution.
We take the magnetic charge $Q_m$ of such monopole as unity. Note, however, that the charges
$Q_m=\pm 1$ are indistinguishable in fact. As for the charges $|Q_m|=2$ they are equivalent, from this
point of view, to no magnetic charge at all.
The $Z_2$ classification might well be illustrated by Wu--Yang monopole
solution \cite{Wu-Yang}. Indeed, the configuration
\beq
\label{Wu-Yang}
A^a_0~=~ 0\,, \qquad
A^a_k~=~ \frac{Q_m}{g} \; \varepsilon^{akn}\; \frac{x_n}{r^2}
\eeq
for $Q_m=1$ represents the (singular) $Z_2$ magnetic monopole , while for $Q_m=2$ it is
gauge equivalent to the vacuum $A=0$.

Within the dynamical, or $U(1)$ classification \cite{U1-classification} one looks for monopole-like
solutions of the classical Yang-Mills equations. Where by the "monopole-like" solutions one understands
potentials which fall off as $1/r$ at large $r$.
The basic finding is that there are no specific non-Abelian solutions and all the monopoles
are, in fact, gauge copies of pure Abelian monopoles embedded into the $SU(2)$ group.
Moreover, using the gauge invariance one can always choose the corresponding $U(1)$ group as, say,
the rotation group around the third direction in the color space.
According to this classification, the monopoles are characterized by their charge with respect
to the $U(1)$ subgroup and may have, therefore, charges,
\beq
\label{quatization-1}
|Q_m|~=~0, ~1,~2,~...
\eeq
Note that the topological solution constructed within $Z_2$ classification exactly
corresponds to the $|Q_m|=1$ monopole in the $U(1)$ approach. For instance,
the $Q_m=1$ configuration (\ref{Wu-Yang}) is gauge equivalent to
\beq
A^a_\varphi ~=~ - \delta^{a,3} \;\frac{1}{g}\, \frac{(1- \cos\theta )}{2r\,\sin\theta}\,,
\qquad
A^a_r = A^a_\theta = 0\,.
\eeq

Therefore we can restrict ourself to the $U(1)$ classification scheme only and
consider every monopole configuration as a pure Abelian one embedded into the
$SU(2)$ group.  The charge quantization condition (\ref{quatization-1}) is
then naturally interpreted as a classification of various admissible Dirac
strings whose end points represent monopoles. Indeed, there is the usual
requirement that the Aharonov--Bohm effect on the Dirac string is not
observable.  It turns out that the Dirac string corresponding to the
$|Q_m|=1$ monopole would be "seen" through the scattering of particles in
the fundamental representation but is invisible to the gluons since their
charges are twice as large.  Contrary to that the $|Q_m|=2$ string
singularities cannot be detected by any test particle.  It is worth
emphasizing that in the standard lattice formulation of the theory the
action is constructed in terms of infinitesimal Wilson loops in the
fundamental representation, which implies in turn that the energy of the
$|Q_m|=1$ Dirac string is infinite in the continuum limit.  Therefore, such
Dirac strings and the corresponding $|Q_m|=1$ monopoles can be introduced
only as external objects via the 't~Hooft loop.

Within the $Z_2$ classification, having the charge $|Q_m|=2$ is equivalent
to have no monopoles at all. This is not true, however, within the $U(1)$
classification.  As is emphasized in Ref.~\cite{coleman}, this apparent
contradiction is resolved through the observation that the Abelian monopoles
with $|Q_m|=2$ are in fact unstable. The instability was demonstrated in
Ref.~\cite{Brandt-Neri} by analyzing the modes of the non-Abelian gluonic
field in the monopole background. Physically the instability corresponds to
the gluon falling onto the monopole.  Contrary to that the monopoles with
$|Q_m|=1$  have no negative modes and are stable in gluodynamics.

Below we briefly describe the $Q_m=2$ monopole instability
following Ref.~\cite{Brandt-Neri}. The quadratic fluctuation operator for the
$Q_m=2$ magnetic monopole has a negative mode corresponding to the $s$--wave
scattering of the charged gauge boson on the monopole. This negative mode
corresponds to the instability for the charged boson to fall onto the
monopole center due to attractive nature of the monopole--boson potential in
the $s$--wave. The operator which determines the stability of the Abelian
monopole solution is just the Sch\"odinger equation for the a charged vector
particle (gluon) in the field of the Abelian monopole~\cite{Brandt-Neri}:

\beqn
\Bigl(\Delta_r \, \delta_{ij} - \frac{1}{r^2} \Lambda_{ij} \Bigr) \,
\psi_j = \omega^2 \, \psi_i\,,
\eeqn
where $\psi_i = \psi_i(r)$ are
spatial components of the spin--one gluon wave function, $\omega^2$ is the
corresponding eigenvalue. The summation over the silent indices
$i,j,k=1,2,3$ is assumed here and below. The operator $\Delta_r = - r^{-1}\,
\partial^2_r \, r$ is the radial part of the three dimensional 
Laplace operator and

\beqn
\Lambda_{ij} = Q_m \, {\mathbf S}_{ij} \cdot
\hat {\mathbf r} - \Bigl({\mathbf L} \cdot {\mathbf L}\, - \frac{Q^2_m}{4}
\Bigr)\, \delta_{ij}\,, \label{eq:lambda}
\eeqn
where ${\mathbf S}_{ij} = i \varepsilon_{ijk} \hat {\mathbf r}_k$ is the
gluon spin operator, ${\mathbf L} = i {\mathbf r} \times ({\mathbf \partial}
+ 2 \pi i Q_m {\mathbf{A}}) - Q_m \hat {\mathbf r} \slash 2$ is the angular
momentum operator, ${\mathbf{A}}$ is the monopole field and $\hat {\mathbf
r} = {\mathbf r} \slash r$. The equations for the angular and radial parts
of the wave function, $\psi_i = X_i(\theta,\phi) R(r)$, are:

\beqn \Bigl(\Delta_r - \frac{1}{r^2} \lambda \Bigr) \,
R_\lambda(r) & = & \omega^2 \, R_\lambda(r)\,, \label{r} \\
\Lambda_{ij} X_j(\theta,\phi) & = & \lambda X_i(\theta,\phi) \, .
\eeqn

For $\lambda \leq 1 \slash 4$ there exists the continuous
spectrum for Eq.\eq{r}, $\omega^2 \geq 0$. This case corresponds to the
scattering of the gluon on the monopole. In this case the Abelian monopole
is stable in gluodynamics. According to Ref.~\cite{Brandt-Neri} this
situation is realized for the $Q_m=1$ monopoles.

If $\lambda > 1 \slash 4$ then there exists infinite number of the monopole-gluon
bound states ($\omega^2 < 0$), for the ground state $\omega^2 =
-\infty$. Physically this situation corresponds to the gluon falling down to
the monopole and it is quite similar to the Callan--Rubakov
catalysis~\cite{callan} of the nucleon decay by monopoles.  The monopole
with charges $Q_m>1$ always contain an eigenvalue $\lambda > 1 \slash 4$ and
thus they are unstable~\cite{Brandt-Neri}.

In a much simpler way, one can demonstrate apparent irrelevance of the
$|Q_m|=2$ monopoles by producing an explicit non-Abelian field configuration
which looks as a $|Q_m|=2$ monopole pair in its Abelian part but has no
$SU(2)$ action at all \cite{main}.  Such a configuration is generated from
the vacuum by the following gauge rotation matrix:
\beq
\label{kyoto-1}
\Omega~=~\left[
\begin{array}{cc}
e^{i\varphi}\sqrt{A_D} &  \sqrt{1-A_D}        \\
-\sqrt{1-A_D}       &  e^{-i\varphi}\sqrt{A_D}
\rule{0mm}{6mm}
\end{array}
\right]\,,
\eeq
where $\varphi$ is the azimuthal angle around the axis connecting the monopoles
and $A_D$ is the $U(1)$ potential representing Abelian monopole pair:
\beq
\label{kyoto-2}
A_{\mu}dx_{\mu}~=~
\frac{1}{2}\;(\frac{z_+}{r_+}-\frac{z_-}{r_-})\; d\varphi ~ \equiv ~ A_D(z,\rho)\; d\varphi\,,
\eeq
$$
z_{\pm}= z\pm R/2\,,\quad \rho^2=x^2+y^2\,, \quad r_{\pm}^2=z_{\pm}^2+\rho^2\,.
$$
Note that the action associated with the Dirac string is considered in this case zero,
in accordance with the lattice version of the theory (for details see \cite{main}).

In this example the monopoles with $|Q_m|=2$ are a kind of pure gauge field configurations
carrying no action at all. Therefore, it seems natural to conclude that only the $Z_2$
classification is relevant. However, the crucial point is that $|Q_m|=2$ monopoles do exist
in the vacuum and, in particular, realize the equivalence of $Q_m=\pm 1$ monopoles
implied by $Z_2$ classification.
Indeed, suppose that we start with, say, $Q_m=+1$ configuration. Then a Dirac
string carrying the $Q_m=-2$ flux can be superimposed on this solution.
It is important at this point that such a Dirac string costs no action (or energy).
Then the radial magnetic field can also change its direction since it does not
contradict the flux conservation any longer. In a related language,
one could say that the $|Q_m|=2$ monopoles are condensed in the
vacuum and that is why the magnetic charge can be changed freely
by two units.
As a result, one expects that two $|Q_m|=1$ monopoles
would interact as a monopole-antimonopole pair, since their
attraction corresponds to the lowest energy state of the system.

Thus, any monopole-like configuration in gluodynamics
may be considered within
the $U(1)$ classification scheme. At short distances, $r \ll \Lambda_{QCD}^{-1}$ only the
fundamental $|Q_m|=1$ monopoles can consistently be constructed.
Moreover, they are infinitely heavy and thus only exist
as external objects. On the other hand, $|Q_m|=2$ monopoles do not produce any action
at small distances, they are so to say empty at this scale. This fact implies in turn
that the $Q_m=\pm 1$ monopoles are indistinguishable from each other and therefore
effectively describe the $Z_2$ monopoles.
Contrary to that, the monopole dynamics at distances $r \gtrsim \Lambda_{QCD}^{-1}$
is much more subtle. In particular, one cannot prove any longer that the action
associated with $|Q_m|=2$ monopoles
at large distances vanishes. We discuss the dynamics of $|Q_m|=2$ monopoles
in Section \ref{Effective-Model}.

\section{Monopoles in the Continuum Limit \\ of SU(2) Lattice Gauge Theory.}
\label{cont-monopoles}

We have shown that all monopoles in the $SU(2)$ gauge theory
may be considered in a unified way within the dynamical $U(1)$ classification.
It is worth emphasizing that the only possibility to have a consistent
description of monopoles is to consider the continuum theory as a limiting
case of lattice formulation. Indeed, we have argued that the classification of
monopoles reduces to the classification of all admissible Dirac strings, while
in the conventional continuum theory the Dirac strings are not allowed at all.
Thus we are encouraged to consider all admissible string-like singularities in the continuum
limit of lattice gauge models.

Following \cite{main} we start from the $SU(2)$ lattice gauge theory (LGT) with the standard Wilson action
\beq
\label{lat-action}
S=-\sum_p S_p \,, \qquad
S_p= \frac{2}{g^2} \tr U_p \,,
\eeq
where $g$ is the bare coupling constant, $p$ is an elementary plaquette and
$U_p$ denotes the ordered product of link matrices in the fundamental representation
along the boundary of $p$.
The matrix $U_p = e^{i F_p}$ is defined in terms of the lattice field strength tensor
$F_p = F^a_p \sigma^a/2$. Therefore, the action (\ref{lat-action}) depends only on
$\cos(|F_p|/2)$, $|F_p| = (F^a_p F^a_p)^{1/2}$
and thus possesses not only the usual gauge symmetry, but allows also
for the gauge transformations which shift the field strength by $4\pi k$,
$|F_p|\to |F_p|+ 4\pi k$, $k\in Z$.
For $|F_p|\ne 0$ we define $n^a_p = F^a_p/|F_p|$ and let $n^a_p$ be an arbitrary
unit vector for $|F_p| = 0$.
Then the symmetry inherent to the lattice formulation may be formulated as:
\beq
\label{lat-sym}
F^a_p ~\to ~ F^a_p ~+~ 4\pi n^a_p\,, \qquad
(n^a_p)^2 ~=~ 1\,, \qquad
\varepsilon^{abc} \, F^b_p \, n^c_p ~=~ 0\,,\;\; a=1,2,3\,.
\eeq
The symmetry (\ref{lat-sym}) which directly follows from the compactness of $SU(2)$,
is evidently lost in the naive continuum limit.
Therefore, the action of $SU(2)$ LGT in the continuum limit is not given by the
conventional expression $\sim \int (F^a_{\mu\nu})^2$ since it should respect
the continuum analog of Eq.~(\ref{lat-sym}).

One can show that in the continuum limit $n^a_p$ represents a singular two-dimensional string
$\Sigma^a_{\mu\nu}$ with a particular color orientation. The explicit construction of
$\Sigma^a_{\mu\nu}$ is as follows (see Ref.~\cite{main} for details). Let $\Sigma_{\mu\nu}$ denote
a non self-intersecting surface with world-sheet coordinates $\tilde{x}(\sigma)$, parameterized by
$\sigma_\alpha$, $\alpha=1,2$:
\beq
\label{Sigma}
\Sigma_{\mu\nu} = \int d^2\sigma_{\mu\nu} \; \delta^{(4)}(x-\tilde{x}(\sigma))\,,
\qquad
d^2\sigma_{\mu\nu} = d^2\sigma \; h_{\mu\nu} \,,
\qquad
h_{\mu\nu} ~=~ \varepsilon^{\alpha\beta} \diff_\alpha \tilde{x}_\mu \diff_\beta \tilde{x}_\nu\,.
\eeq
For given $\Sigma_{\mu\nu}$ consider the value of the dual field strength tensor
$\dual{F}^a_{\mu\nu} = \frac{1}{2} \varepsilon_{\mu\nu\lambda\rho} F^a_{\lambda\rho}$
tangent to the surface:
\beq
\dual{F}^a(\sigma) ~=~ \dual{F}^a_{\mu\nu} \; h_{\mu\nu}\,,
\qquad
F^a_{\mu\nu} = \diff_{[\mu} A^a_{\nu ]} + g \, \varepsilon^{abc} A^b_\mu A^c_\nu\,.
\eeq
In close analogy with the lattice considerations above we define:
\beq
\label{const-n-a}
n^a(\sigma) ~=~ \frac{\dual{F}^a}{|\dual{F}|} \qquad \mbox{for} \qquad
|\dual{F}|^2 = (\dual{F}^a)^2  \ne 0\,,
\eeq
while $n^a$ is an arbitrary unit vector for all points in which $|\dual{F}|=0$. Then the
surface $\Sigma^a_{\mu\nu}$ is constructed as follows:
\beq
\label{Sigma-colored}
\Sigma^a_{\mu\nu} = \int d^2\sigma_{\mu\nu}\; n^a \; \delta^{(4)}(x-\tilde{x}(\sigma))\,,
\eeq
where we have used the notations (\ref{Sigma}). With this definition
the lattice symmetry relation (\ref{lat-sym}) becomes in the continuum limit:
\beq
\label{con-sym}
F^a_{\mu\nu} ~ \to ~ F^a_{\mu\nu} ~+~ \frac{4\pi}{g} \dual{\Sigma}^a_{\mu\nu}\,.
\eeq

In order to construct a continuum action which respects Eq.~(\ref{con-sym}), note that
this relation should be valid for any shape of $\Sigma$, Eq.~(\ref{Sigma}). On the other hand,
in the sector with no surfaces at all this action should reduce to the standard one
$1/4 \int (F^a_{\mu\nu})^2$. The most straightforward approach would be to sum
over all possible strings $\Sigma$:
\beq
\label{con-action}
S(F) ~=~  -\ln \; \int \cD \Sigma \exp\{\;
-\frac{1}{4} \int \mathrm{d}^4 x \; (F^a_{\mu\nu} ~+~ \frac{4\pi}{g} \dual{\Sigma}^a_{\mu\nu})^2 \;\}\,.
\eeq
Note that the Eq.~(\ref{con-action}) is much analogous to the Villain formulation of
the compact electrodynamics.
The corresponding partition function of the $SU(2)$ gluodynamics becomes therefore:
\beq
\label{con-PF}
Z ~=~ \int \cD A \exp\{\;-S(F)\;\} ~=~
\int \cD A \cD \Sigma \exp\{\;
-\frac{1}{4} \int \mathrm{d}^4 x \; (F^a_{\mu\nu} ~+~ \frac{4\pi}{g} \dual{\Sigma}^a_{\mu\nu})^2 \;\}\,.
\eeq

To appreciate the meaning of the integral $\int \cD\Sigma$ it is sufficient to consider one
particular surface in Eq.~(\ref{con-PF}). It turns out that $\Sigma^a$ exactly
corresponds to the Dirac string associated with $|Q_m|=2$ monopole. Thus, the integration
$\int \cD\Sigma$ implies invisibility of the $|Q_m|=2$ Dirac strings
in the continuum limit of $SU(2)$ LGT. The appearance of the additional color index is due
to the Abelian nature of monopoles, which might have an arbitrary color orientation
within $SU(2)$. In particular, for non self-intersecting surfaces it is always
possible to find a gauge in which all the Dirac strings are aligned along, say, third
direction in the color space $\Sigma^a = \delta^{a,3}\Sigma$.

Note also that in Eqs.~(\ref{con-action}, \ref{con-PF}) the integral $\int \cD\Sigma$
has, in a sense, a symbolic meaning: it is impossible to separate rigorously the measure
$\cD\Sigma$ from the gauge degrees of freedom in $\cD A$.  A concrete example
of this kind is provided by the Eqs.~(\ref{kyoto-1},\ref{kyoto-2}) which show that the
string singularities might be generated by gauge transformations.

It is also possible to derive an analogous representation for $|Q_m|=1$ monopoles.
We have already noted that these monopoles are not inherent to the vacuum of gluodynamics
and may only be introduced as an external objects via 't~Hooft loop. Consider therefore
the lattice definition of the 't~Hooft loop operator:
\beq
\label{tHooft-loop-lattice}
H_{lat}(\Sigma_j) ~=~ \exp\{ - 2  \sum\limits_{p \in \dual{\Sigma}_j} S_p \}\, ,
\eeq
where $\Sigma_j$ is an arbitrary surface spanned on a given contour $j$ (the star symbol
means the duality operation). In the path integral formulation the 't~Hooft loop effectively
changes the sign of the plaquette variables $U_p$ belonging to $\dual{\Sigma}_j$ : $U_p \to - U_p$,
which can also be represented as:
\beq
\label{tHooft-temp}
F^a_p ~\to ~ F^a_p ~+~ 2\pi n^a_p \qquad \mbox{for} \qquad p\in\dual{\Sigma}_j\,,
\eeq
where $n^a_p$ is defined exactly as before. The Eq.~(\ref{tHooft-temp}) is almost identical
to Eq.~(\ref{lat-sym}) and therefore one can appreciate the continuum limit of (\ref{tHooft-loop-lattice}).
Namely, the 't~Hooft loop operator which introduces the $|Q_m|=1$ monopoles into the vacuum of $SU(2)$
gluodynamics is given by~\cite{main}:
\beq
\label{tHooft-loop-operator}
H(\Sigma_j) ~=~ \exp\{ ~S(F) ~-~ S(F + \frac{2\pi}{g} \dual{\Sigma_j})~\}\,,
\eeq
where the action $S(F)$ is defined in Eq.~(\ref{con-action}).
Note that if there was no explicit integration $\int \cD\Sigma$ in (\ref{con-action}) then the operator
(\ref{tHooft-loop-operator}) would insert the $U(1)$ monopole pair in which the $Q_m=1$
and $Q_m=-1$ particles are different. This would be in clear contradiction with the general
properties of the 't~Hooft loop. Thus it is precisely the $|Q_m|=2$ monopoles
which ensure the $Z_2$ nature of the operator (\ref{tHooft-loop-operator}).

\section{$Z_2$ monopoles at short distances}
\label{Z_2-short-distances}

It is crucial for understanding the symmetries of the problem that the $|Q_m|=1$
monopoles are point-like objects. Therefore, we are challenged
to consider their interaction at arbitrary small distances or, in other words,
at the fundamental level. Moreover, the monopoles in gluodynamics
are in fact Abelian. Therefore, we immediately come to the paradoxical conclusion that the dual field,
if any, is Abelian as well. There is no place for a non-Abelian dual gluon because
the monopoles do not constitute representations of the non-Abelian group. Then
an immediate problem is
how to match the non-Abelian symmetry of
the gluon interactions with the color and the Abelian symmetry of the interaction of the
dual gluon with the external magnetic charges.
We will discuss the problem in detail in this section.
Basically, our approach is similar to that of Zwanziger \cite{zwanziger}
in case of electrodynamics. However, before going into this subject let us consider in more
detail the expectation value of the operator (\ref{tHooft-loop-operator}).

Here, we will restrict ourselves to the rectangular $T\times R$,
$T \gg R$ contours $j$. Then
\beq
\label{H-expectation}
H(j) ~\equiv ~\langle H(\Sigma_j) \rangle ~\sim ~ e^{-T\; V_{m\bar{m}}(R)}
\eeq
and we refer to $V_{m\bar{m}}(R)$ as the inter-monopole
(mo\-no\-po\-le--anti\-mo\-no\-po\-le) potential.
For simplicity, we consider only the sector with no $|Q_m|=2$ monopoles in which
the action (\ref{con-action}) coincides with the standard one and therefore:
\beq
\label{prem}
H(j)~=~ \frac{1}{Z} \int \cD A \; \exp\{\;
- \frac{1}{4} \int \mathrm{d}^4 x \; (F^a_{\mu\nu} ~+~ \frac{2\pi}{g} \,\dual{\Sigma}^a_{j\;\mu\nu})^2
\;\}\,.
\eeq
It is easy to see that for the contours considered the potential $V_{m\bar{m}}(R)$ calculated
with (\ref{prem}) coincides with the potential of a $Z_2$ monopole pair.

To derive $V_{m\bar{m}}(R)$ at the classical level, note
first that the gauge covariance of Eq.~(\ref{prem}) allows to consider
a particular gauge in which $\Sigma^a_j=\delta^{a,3}\Sigma_j$. Then the classical solution of the
field equations corresponding to (\ref{prem}) is:
\beq
\label{small-R-solution}
\tilde{A}^3_\mu dx_\mu ~\equiv ~ \tilde{A}_\mu dx_\mu
= {1\over 2g}\left( {z_+ \over r_+} - {z_- \over r_-}\right) \; d\varphi\,,
\qquad
\tilde{A}^{1,2}_\mu = 0\,,
\eeq
$$
z_\pm  = z \pm R/2\,, \qquad \rho^2 = x^2 + y^2\,, \qquad r^2_\pm = z^2_\pm + \rho^2\,.
$$
Moreover, it can be shown \cite{main,rubakov} that the general classical solution with
minimal energy is a gauge rotation of (\ref{small-R-solution}). Thus the classical inter-monopole
potential is:
\beq
\label{small-R-potential}
V_{m\bar{m}}(R) ~=~ -{\pi \over g^2\;R}\,.
\eeq
It is worth emphasizing that the classical limit of the state created by the non-Abelian 't~Hooft loop
operator is an Abelian (up to the gauge transformation) mo\-no\-po\-le-anti\-mo\-no\-po\-le pair.

Next, consider radiative corrections to the Coulomb-like interaction (\ref{small-R-potential})
at short distances. There are two specific problems
which have to be addressed. First, the classical solution
(\ref{small-R-solution}) has
a Dirac-string singularity. Second, the radial magnetic field is
also inversely proportional to the
coupling constant and at first sight this mixes up all orders of the perturbation theory.

To sort the things out, note
that the evaluation of radiative corrections addresses in fact
two different, although closely related problems. These are the running of the coupling and stability
of the classical solutions. Both aspects are unified, of course, into evaluation of a single loop
in the classical background. However, the running of the coupling can be clarified by keeping track
of the ultraviolet logs, $\ln\Lambda^2_{UV}$ alone and is universal since in the ultraviolet all the
external fields can be neglected. Therefore, the coefficient in front of $\ln \Lambda_{UV}$ can be
found by evaluating the loop graph with two external legs, i.e. the graph corresponding to the standard
polarization operator in perturbation theory. This is true despite of the fact that the
monopole field is of order $1/g$. On the other hand the stability of the classical solution is
decided by the physics in the infrared. Here one needs to consider the particular dynamical system,
monopoles in our case, and the fact that the magnetic charge is of order $1/g$ can be crucial.

Thus to derive the running of the coupling we have to consider only the terms quadratic in the
external field and demonstrate that indeed the leading quantum corrections produce
$g^2 \to g^2(R)$. However, we have ignored so far the string singularity, which in fact invalidates
direct perturbative calculations. Nevertheless, we argue that one can still use the perturbation theory
supplemented by an additional rules, which are to remove the effect of Dirac strings so to say
by hand. These additional rules and their meaning are  discussed in detail in subsection 4.2.

Consider the standard decomposition $A_\mu= \tilde{A}_\mu + a_\mu$ of the gauge potentials $A_\mu$
into classical field $\tilde{A}_\mu$, Eq.~(\ref{small-R-solution}), and quantum fluctuations $a_\mu$
taken in the background gauge $\tilde{D}_\mu a_\mu = 0$.
Since the color orientation of $\Sigma^a_j$ is a functional of gauge potentials,
in account of quantum perturbations the vector $n^a$ becomes
\beq
\label{n-deviates}
n ~=~ n_{cl.} ~+~ \delta n \,, \qquad
\delta n^a ~=~ \varepsilon^{abc} \; n^b_{cl.} \; \omega^c\,,
\eeq
where the small parameter $\omega$ is of order $a_\mu$. But in fact one can neglect the deviation
$\delta n$ and consider $n = n_{cl.}$ instead (similar arguments may be found in Ref.~\cite{Brandt-Neri}).
Indeed, the condition $\tilde{D}_\mu a_\mu = 0$
is invariant under $a_\mu\to a_\mu+\tilde{D}_\mu\lambda$ provided that the gauge parameter $\lambda$
satisfies $\tilde{D}^2 \lambda =0$. The second order differential equation  $\tilde{D}^2\lambda = 0$
supplemented by the boundary condition  $\lambda = -\omega $ on the surface $\Sigma_j$ always has
a solution. Since $n$ transforms in the adjoint representation the gauge transformation
$a_\mu\to a_\mu+\tilde{D}_\mu\lambda$ brings the $n^a$ to its classical value
$n^a_{cl.}= \delta^{a,3}$.

Then the effect of one-loop corrections to the potential (\ref{small-R-potential}) reduces
to the calculation of
\beq
\label{det-1}
\mathrm{Det}^2[\, -\tilde{\cD}^2 \,]\cdot
\mathrm{Det}^{-1}[\,-\tilde{\cD}^2 \,\delta_{\mu\nu} + 2i g \tilde{f}_{\mu\nu}\,]\,,
\eeq
where the first term is due to the ghosts, $\tilde{\cD}_\mu = \diff_\mu -ig \tilde{A}_\mu$
is a $U(1)$ covariant derivative  and $\tilde{f}_{\mu\nu}$ denotes the regular part of the classical field:
\beq
F^a_{\mu\nu}(\tilde{A}) ~=~
\delta^{a,3} \, (
- \frac{2\pi}{g} \; \dual{\Sigma}_{j\;\mu\nu} ~+~ \tilde{f}_{\mu\nu})\,.
\eeq
In the leading log approximation the second term in Eq.~(\ref{det-1}) factories:
\beq
\mathrm{Det}^{-1}[\,-\tilde{\cD}^2 \,\delta_{\mu\nu} + 2i g \tilde{f}_{\mu\nu}\,] ~=~
\mathrm{Det}^{-4}[\, -\tilde{\cD}^2 \, ]\cdot \mathrm{Det}^{-1}[\,2i g \tilde{f}_{\mu\nu}\,]\,.
\eeq
Thus the paramagnetic interaction is free of string singularity. On the other hand,
the diamagnetic interaction term apparently gives the contribution proportional to
$\int ( \diff_{[\mu} \tilde{A}_{\nu]} )^2$, which includes also the string singularity.
To decide whether the Dirac string becomes "visible" at the level of quantum corrections
one can consider the evaluation of $\mathrm{Det}[-\tilde{\cD}^2]$  in the field of single monopole \cite{goebel},
where the determinant may be calculated exactly. It turns out that
$\mathrm{Det}[-\tilde{\cD}^2]$ is proportional to the squared regular magnetic field,
$\int (\tilde{f}_{\mu\nu})^2$ and is string independent.

Thus the perturbative and exact calculations of the leading $\ln\Lambda_{UV}$ behavior
lead to the distinct results, which differ however by the Dirac string contribution.
The reason for this difference is not difficult to figure out.
Indeed, in the theory of charged particles
interacting with monopoles there is a famous Dirac "veto" which forbids any direct interaction
with string singularity. But exactly this prohibition is violated in ordinary perturbation theory,
which uses the plane-wave basis in calculations. This is in sharp contrast with exact evaluation
of $\mathrm{Det}[-\tilde{\cD}^2]$, where one uses the basis of eigenfunctions which explicitly
depend on given monopole background.
The only way to make sense of perturbative result is to substitute
$\diff_{[\mu} \tilde{A}_{\nu]} \to \tilde{f}_{\mu\nu}$, thus discarding the string contribution.
Evidently, this substitution goes beyond the field theory and should be considered as an
additional rules in perturbative calculations.
We will return to the discussion of this issue in section \ref{Z_2-2}.

The analytical calculation of $\mathrm{Det}[-\tilde{\cD}^2]$
in the case of monopole-anti\-mo\-no\-po\-le pair is impossible to perform \cite{goebel}.
Nevertheless, in discussing the string independence the problem is in fact identical to
the single monopole case. Thus the conclusion is that $\mathrm{Det}[-\tilde{\cD}^2]$
depends only on the regular part of $F^a_{\mu\nu}(\tilde{A})$ from which the result
for the inter-monopole potential at the one-loop level follows:
\beq
\label{q-pot}
V_{m\bar{m}}(R) ~=~ -{\pi \over g^2(R)\;R}\,.
\eeq

In the next section we consider the expectation value  (\ref{H-expectation},\ref{prem}) anew
and introduce the dual representation for $H(j)$.
To test the validity of new formulation we calculate in section \ref{Z_2-2}
the monopole-antimonopole potential up to the one-loop order and reproduce the running of
the coupling (\ref{q-pot}).

\subsection{Lagrangian Approach}
\label{Z_2-1}

In this section we derive a kind of dual representation for the expectation value
(\ref{H-expectation},\ref{prem}). Since our approach is close to that of
Zwanziger~\cite{zwanziger}, it is worth to highlight the essentials of the latter construction.

The Zwanziger Lagrangian which describes interaction of a $U(1)$ gauge fields with
Abelian point-like monopoles looks as:
\beq
\label{Zw-action}
L_{\mathrm{Zw}}(A,B)~=~
\frac{1}{2}(m\cdot[\diff\wedge A])^2 ~+~ \frac{1}{2}(m\cdot[\diff\wedge B])^2 ~+
\eeq
$$
+~\frac{i}{2}(m\cdot[\diff\wedge A])(m\cdot\dual{[\diff\wedge B]}) ~-~
\frac{i}{2}(m\cdot[\diff\wedge B])(m\cdot\dual{[\diff\wedge A]})
~+~ i\,g_e\,j^e\cdot A ~+~ i\,g_m\,j^m\cdot B\,,
$$
where $j^e$ ($j^m$) and $g_e$ ($g_m$)  are electric (magnetic) currents and coupling, respectively,
$m_{\mu}$ is a constant vector, $m^2=1$ and
\beq
[A\wedge B]_{\mu\nu} = A_\mu B_\nu - A_\nu B_\mu\,, \qquad
(m \cdot [A\wedge B])_\mu = m_\nu [A\wedge B]_{\mu\nu}\,.
\eeq
At first sight, there are two different vector fields, $A$, $B$ introduced to describe
interaction with electric and magnetic charges. If it were so, however,
a wrong problem would have been solved because we need to have a single photon interacting
both with electric and magnetic charges. And this is what is achieved by the construct
(\ref{Zw-action}). Indeed, the action (\ref{Zw-action}) is not diagonal
in the $A$, $B$ fields and one can check that the form of the $A,B$
interference terms in (\ref{Zw-action}) is such that the field strength tensor constructed
on the potentials $A$ and $B$ are in fact related to each other.
To be more specific, let us consider $A$ and $B$ potentials which for example are due to
a single magnetic current $j^m$. Using the propagators (\ref{propagators}) to be derived
shortly and the relation
$G + [m\wedge (m\cdot G)] + \dual{[m\wedge (m\cdot \dual{G})]} = 0$
valid for any antisymmetric tensor $G_{\mu\nu}$ one can show that
\beq
\label{Zw-constraint}
[\diff\wedge A] + i \dual{[\diff\wedge B]} = -i (m\diff)^{-1} \dual{[m\wedge j^m]}\,.
\eeq
The right hand side of this equation is non-zero only on the Dirac string
attached to the current $j^m$ and directed along $m_\mu$. Thus the constraint (\ref{Zw-constraint})
implies that there are only two physical degrees of freedom corresponding to the transverse
photons which can be described either in terms of the potential $A$ or $B$.
Topological excitations, however, can be different in terms of $A$ and $B$.

The physical content of (\ref{Zw-action}) is revealed by the propagators of the fields $A,B$.
Since the Lagrangian (\ref{Zw-action}) is at most quadratic in $A$, $B$ it is easy to integrate
out potentials $A,B$ and obtain the corresponding propagators. In the Feynman
gauge they are
\beq
\label{propagators}
\langle A_{\mu} A_{\nu}\rangle ~=~ \langle B_{\mu} B_{\nu}\rangle ~=~ \frac{\delta_{\mu\nu}}{k^2}\,,
\eeq
$$
\langle A_{\mu} B_{\nu}\rangle ~=~ - \langle B_{\mu} A_{\nu}\rangle ~=~
\frac{i}{k^2 (km)} \; \dual{[m\wedge k]}_{\mu\nu}\,.
$$
The propagators should reproduce, as usual, the classical solutions. And indeed,
the $\langle AA \rangle$, $\langle BB \rangle$ propagators describe
the Coulomb-like interaction of two charges and magnetic monopoles, respectively.
While the $\langle AB \rangle$ propagator reproduces interaction of the magnetic field
of a monopole with a moving electric charge.

Consider now the expectation value (\ref{H-expectation},\ref{prem}). Since it depends only
on the current $j$ and not on the particular shape of $\Sigma_j$ one can take
\beq
\label{Sig-1}
\Sigma_j ~=~ \frac{1}{(m\diff)} [m \wedge j]\,,
\eeq
which is a particular solution to $\partial \Sigma_j=j$.
For rectangular contours $j$ which we consider the surface (\ref{Sig-1}) has no self-intersection
points. Therefore one can write
\beq
\label{Sig-col-1}
\Sigma^a_j ~=~ n^a \frac{1}{(m\diff)} [m \wedge j]
\eeq
thus extending the definition of the field $n^a$ into the
entire space--time. Note that only the value
of $n^a$ on the string $\Sigma_j$ does matter
while a particular way of extending $n^a(\sigma) \to n^a(x)$
is irrelevant. Then the expectation value $H(j)$ is represented as:
\beq
\label{H-2}
H(j) ~=~ \frac{1}{Z}\int \cD A \int\limits_{n^2=1} \!\!\! \cD n \;
\exp\{\; - \frac{1}{4} \int \mathrm{d}^4x \, (
F^a + \frac{2\pi}{g} \,n^a\, \frac{1}{(m\diff)} \dual{[m\wedge j ]} )^2 \;\}\,,
\eeq
where the path integral is to be performed  with a constraint
\beq
\label{constr}
\varepsilon^{abc} n^b ( F^c \cdot \frac{1}{(m\diff)} \dual{[m\wedge j]})~=~0 , \qquad a=1,2,3\,,
\eeq
which is a consequence of Eq.~(\ref{const-n-a}).
The constraint equation (\ref{constr}) may be implemented by an additional field $\chi^a$:
\beq
\label{H-3}
H(j)= \frac{1}{Z}\int\limits_{n^2=1} \!\!\! \cD n \cD A \cD \chi \,
\exp\{\; - \frac{1}{4} \int \! \mathrm{d}^4x \,
(F^a + \frac{2\pi}{g} \,n^a\, \frac{1}{(m\diff)} \dual{[m\wedge j ]} )^2
+ i\frac{2\pi}{g}\!\!\int\! j C \;\}\,,
\eeq
\beq
\label{C}
C_\mu ~=~ C_\mu(\chi, n, F )~=~
\frac{1}{(m\diff)} \varepsilon^{abc} \chi^a n^b ( m \cdot \dual{F}^c )_\mu\,.
\eeq
It is worth emphasizing that the field $n^a(x)$ is a kind of a fake variable in the
representation (\ref{H-3}). The path integral is clearly independent on $n^a$ away from
the string (\ref{Sig-1}), but at the same time $n^a(x)$ for $x\in \Sigma_j$
is determined through (\ref{constr}).

The last step in deriving the dual formulation of (\ref{H-expectation},\ref{prem})
is to use the equality:
\beq
\label{B-1}
\exp\{\; -\frac{1}{4} \int \! \mathrm{d}^4x \, (G + \frac{1}{(m\diff)} \dual{[m\wedge j]})^2 \;\} ~=~
\int \cD B \exp\{\; - \int \! \mathrm{d}^4x \, L(G,B,j) \;\} \,,
\eeq
$$
L(G,B,j) ~=~ \frac{1}{4} G^2 ~+~  \frac{1}{2}[m\cdot( \diff\wedge B  ~-~ i\dual{G} ) ]^2 ~+~ i\, jB \,,
$$
which is valid for any antisymmetric tensor field $G_{\mu\nu}$. The relation
$(m\cdot G)^2 + (m\cdot\dual{G})^2 = G^2 /2$ has been used to derive (\ref{B-1}).
Using Eq.~(\ref{B-1}) and redefining $B \to B - C$
we can finally transform (\ref{H-2}) into the form:
\beq
\label{dual-H}
H(j)~=~\frac{1}{Z}\int \cD A \cD B \cD \chi \! \int\limits_{n^2=1} \!\!\! \cD n \;
\exp\{\; -\int L(A,B,\chi,n) ~+~ i \frac{2\pi}{g} \int \! j\,B  \;\}\,,
\eeq
where Lagrangian $L$ depends on the gauge potentials $A^a_\mu$, Abelian vector field $B_\mu$ and
is independent on the current $j$:
\beq
\label{dual-L}
L ~=~ \frac{1}{4}( F^a_{\mu\nu})^2 ~+~
\frac{1}{2} \Bigr[ 
	\Bigr(
		m\cdot [ \diff\wedge B  ~-~
		i\;\dual{F^a n^a} ~-~
		\varepsilon^{abc} \chi^a n^b \dual{F}^c \, ]
	\Bigr)_\mu\,
\Bigr]^2
\eeq
Let us discuss the properties of the formulation (\ref{dual-H})-(\ref{dual-L}).

In absence of the external magnetic currents $j$ the Lagrangian (\ref{dual-L})
reduces to the standard Lagrangian of $SU(2)$ gluodynamics, as it should be.
Indeed, for vanishing $j$ the integration over $B$ field cancels
the last term in Eq.~(\ref{dual-L}) regardless of $n^a$ (cf. Eq.~(\ref{B-1})).

The Lagrangian (\ref{dual-L}) as well as the interaction terms in Eq.~(\ref{dual-H})
obviously possesses $SU(2) \times U(1)$ gauge invariance:
\beq
\label{freedom}
SU(2)\,:\;
	\begin{array}{ccc}
		F_{\mu\nu} & \to &  \Omega^{-1}  F_{\mu\nu}  \Omega  \\
		n^a \sigma^a \,=\, n & \to & \Omega^{-1}  n  \Omega  \rule{0mm}{5mm} \\
		\chi^a \sigma^a \,=\, \chi & \to & \Omega^{-1}  \chi  \Omega \rule{0mm}{5mm} \\
	\end{array}\,,
\qquad
U(1)\,:\;
	\begin{array}{c}
		B_\mu  ~\to ~ B_\mu + \diff_\mu \alpha \\
	\end{array}\,.
\eeq
As is evident from the transformation law (\ref{freedom}) the expectation value (\ref{dual-H}) may also
be written as
\beq
\label{H_n}
H(j)~=~ \int\limits_{n^2=1} \!\!\! \cD n \;   H_{n}(j)\,,
\eeq
where $H_{n}(j)$ is given analogously to Eq.~(\ref{dual-H}) but with a fixed $n^a$. Thus $H(j)$
is a gauge singlet part of an apparently gauge variant expression $H_n(j)$.
Moreover, the integration $n^a$ is in fact
redundant: the operator $H_n(j)$ is well defined for
any choice of $n^a$ and coincides with $H(j)$. This property of the 't~Hooft loop is well known
\cite{tHooft-loop}. In particular, in the Hamiltonian formulation the 't~Hooft loop operator
may also be defined in a gauge non-invariant way \cite{korthals}. The matrix elements of this operator
taken between physical and unphysical states vanish and this
provides a well defined meaning to the 't~Hooft loop despite of gauge variance.
The same effect may clearly be seen in the formulation (\ref{H_n}). Indeed, since various choices of
$n^a$ are related by gauge transformations one can show that $H_{n_1}(j) = H_{n_2}(j) = H(j)$.
Note also that for every particular $n^a$ one still has to enforce the constraint  (\ref{constr}),
which for $n^a=\delta^{a,3}$ becomes:
\beq
\label{F-on-Sigma}
(\; F^a \cdot \frac{1}{(m\diff)} \dual{[m\wedge j]}\;)~=~0\,, \qquad a=1,2\,.
\eeq

It is evident that the origin of the vector $n^a$ goes
back to the freedom of choosing the color orientation of the monopoles.
As is emphasized above the monopole solutions are Abelian in nature which means, in particular,
that they can be gauge rotated to any direction in the color space. Thus, picking up a
particular $n^a$ is nothing else but choosing a $U(1)$ subgroup to which the
monopoles belong. Note that the embedding of relevant $U(1)$, which is described by $n^a(x)$,
may also be space-time dependent. In particular, the requirement (\ref{F-on-Sigma}) means that
all the monopoles are in the diagonal subgroup of $SU(2)$.
We can either average over the directions of $n^a$ or fix it arbitrarily but evaluate only
gauge invariant quantities.

As far as the quantization is concerned, the Lagrangian (\ref{dual-L}) reproduces the correct
degrees of freedom of the free gluons. Indeed, in the limit $g\to 0$ and for $n^a=\delta^{a,3}$
the Lagrangian (\ref{dual-L}) reduces to
\beq
\label{dual-L0}
L^0 ~=~ \frac{1}{4}[\diff\wedge A^a]^2 ~+~
\frac{1}{2}(m\cdot [ \diff\wedge B  ~-~ i\;\dual{\diff\wedge A^3}])^2\,,
\eeq
which essentially coincides with that of Zwanziger (\ref{Zw-action}) plus free Lagrangian
for off-diagonal gluons. Quantization at this point is the same as in the case of a single
photon. In particular, to the lowest order in $g^2$ the monopole-antimonopole interaction
is given by a single $B$-field exchange. Using Eq.~(\ref{propagators}) one immediately
reproduces the result (\ref{small-R-potential}):
\beq
H(j) ~=~ \exp\{\;  - \frac{1}{2} \, (\frac{2\pi}{g} )^2 \, \int j_\mu \Delta^{-1} j_\mu \;\} ~\sim ~
\exp\{\;  T \frac{4\pi^2}{g^2} \Delta^{-1}_{(3)}(R)\;\}
\eeq
\beq
\label{tree}
V_{m\bar{m}}(R) ~=~ -\frac{1}{g^2}\cdot\frac{\pi}{R}
\eeq

An apparent application of (\ref{dual-L}) would be evaluating the running of the coupling $g$ in
the expression (\ref{tree}). And, indeed, exploiting the Lagrangian (\ref{dual-L}) one can approach
this problem in a way quite similar to the case of pure electrodynamics,
for a review and further references see \cite{blagoevic}. We consider the perturbative corrections
to the potential (\ref{tree}) in the next section.

\subsection{Radiative corrections.}
\label{Z_2-2}
In this section we consider the radiative corrections to the potential (\ref{tree})
starting from (\ref{dual-H}) with $n^a=\delta^{a,3}$. Since for a constant $n^a$
the problem is quite similar to the consideration of Dirac monopole in the electrodynamics,
there is not much specific about the derivation of the running of the coupling. And, indeed, our
considerations overlap to some extent with those given in
the original papers \cite{goebel,eg,calucci} and in the reviews
\cite{blagoevic,coleman}. Still, we feel that it is useful to present
the arguments, may be in a new sequence, to emphasize the points crucial for
our purposes.

Therefore we first concentrate on the Dirac monopole with the minimal magnetic charge interacting
with electrons and in one-loop approximation \cite{blagoevic,goebel,eg,calucci}. The crucial
point which has been already
mentioned above is that only loops with insertion of two external
(i.e., monopole) fields can be considered despite of the fact that there is no perturbative
expansion at all. Indeed, considering more insertions makes the graphs infrared sensitive,
with no possibility for $\ln\Lambda_{UV}$ to emerge.

Then, the evaluation of, say, first radiative correction to the propagator
$\langle B_{\mu}B_{\nu}\rangle$ in the Zwanziger formalism  (\ref{propagators})
seems very straightforward and reduces to taking a product of two
$\langle AB \rangle$ propagators and inserting in between the standard polarization operator
of two electromagnetic currents.  The result is \cite{eg}:
\beq
\label{sch}
\langle B_{\mu}B_{\nu}\rangle(k) ~=~{\delta_{\mu\nu}\over k^2} (1-L)
~+~ {1\over (k\cdot m)^2}(\delta_{\mu\nu}-m_{\mu}m_{\nu}) L,
\eeq
$$
L~=~{\alpha_{el}\over 6}\ln{\Lambda_{UV}^2/k^2}\,,
$$
where we have neglected the electron mass so that the infrared cut-off is provided, in the
logarithmic approximation, by the momentum $k$.

At first sight, there is nothing disturbing about the result (\ref{sch}). Indeed, we have a
renormalization of the original propagator which is to be absorbed into the running coupling,
and a new structure with the factor $(k\cdot m)^{-2}$ in front
which is non-vanishing, however, only
on the Dirac string. The latter term would correspond to the renormalization of the Dirac-string
self-energy which we do not follow in any case since it is included into self-energy of the
external monopoles. What is, actually, disturbing is that according (\ref{sch})
the magnetic coupling would run exactly the same as the electric charge,
\beq
\langle A_{\mu}A_{\nu}\rangle(k) ~=~ (1-L){\delta_{\mu\nu}\over k^2}\,,
\eeq
violating the Dirac quantization condition.

The origin of the trouble is not difficult to figure out. Indeed, using the propagator
$\langle AB \rangle$ while evaluating the radiative corrections is equivalent, of course,
to using the full potential corresponding to the Dirac monopole $A_D^{cl}$. Then, switching on the
interaction with electrons would bring terms like  $A_D^{cl}\, \bar{\psi} \gamma\psi$. Since $A_D^{cl}$
includes the potential of the string electrons do interact with the Dirac sheet and we are
violating the Dirac veto which forbids any direct interaction with the string.

Let us demonstrate quantitatively that, indeed,
it is the interaction with the Dirac string that changes the sign of
the radiative correction. This can be done in fact in an amusingly simple way. First, let us note
that it is much simpler to remove the string if one works in terms of the field strength tensor,
not the potential. Indeed, we have ${\bf H} = {\bf H}_{rad} + {\bf H}_{string}$ while in terms of
the potential $A_\mu$ any separation of the string would be ambiguous
(see, e.g. Eq.~(\ref{small-R-solution})).

Thus, we start with relating the potential, or energy to the interference term in the ${\bf H}^2$
field:
\beq
V_{m\bar{m}}~=~ \int {\bf H}_1\cdot {\bf H}_2 \, \mathrm{d}^3 r\,.
\eeq
Now, it is not absolutely trivial, how we should understand the product ${\bf H}_1\cdot {\bf H}_2$.
Consider for example the magnetic field of a single monopole:
\beq
{\bf H}~=~ {\bf H}_{rad} ~+~ {\bf H}_{string}\,, \qquad \mathrm{div~} {\bf H}~=~ 0\,,
\eeq
$$
{\bf H}_{rad} ~=~ - \frac{1}{4\pi}
\mathrm{~grad~}(\frac{1}{|{\bf r}-{\bf r}_1|})\,, \qquad
\mathrm{div~} {\bf H}_{string}~=~ \delta({\bf r}-{\bf r}_1)\,.
$$
Then, by the analogy with the case of two electric charges, we would like to have
the following expression for the interaction energy:
\beq
\label{right}
V_{m\bar{m}} ~=~ \int {\bf H}_{1,rad}{\bf H}_{2,rad}  \,\mathrm{d}^3 r~=~
- \frac{1}{4\pi}\,\frac{1}{|{\bf r}_1-{\bf r}_2|} \,.
\eeq
Note, however, that if we substitute the sum of the radial and string fields
for ${\bf H}_{1,2\;}$, then we would have an extra term in the interaction energy:
\beq
\label{wrong}
\tilde{V}_{m\bar{m}} ~=~
\int \left({\bf H}_{1,rad}{\bf H}_{2,string}+{\bf H}_{1,string}{\bf H}_{2,rad}
\right)\,\mathrm{d}^3 r  ~=~
+ 2\,\frac{1}{4\pi}\,\frac{1}{|{\bf r}_1-{\bf r}_2|} \,.
\eeq
In other words, the account of the string field would flip the sign of the interaction energy!
This contribution, although looks absolutely finite, is of course a manifestation of the singular
nature of the magnetic field ${\bf H}_{string}$ . Note
that the integral in (\ref{wrong}) does not depend on
the shape of the string.

Thus, in the zero, or classical approximation we should write:
\beq
{\bf H}_1\cdot {\bf H}_2~\equiv~ {\bf H}_{1,rad}\cdot {\bf H}_{2,rad}\,.
\eeq
However, in the above example the electrons interact with full potential $A^{cl}_D$ and
therefore the first radiative correction would bring the product of the total
${\bf H}_1\cdot {\bf H}_2$ which includes also the string contribution\footnote{
At this point we assume in fact that $\Lambda_{UV}$ is smaller than the inverse size of the string,
which is convenient for our purposes. Other limiting procedures could be considered as well,
however.
}. Indeed, the result in the log approximation
would be as follows:
\beq
\label{delta}
\delta ({\bf H}_1\cdot {\bf H}_2)~=~L({\bf H}_{1,string}+{\bf H}_{1,rad})\cdot
({\bf H}_{2,string}+{\bf H}_{2,rad})~=~-L {\bf H}_{1,rad}\cdot {\bf H}_{2,rad}\,,
\eeq
where at the last step we have used the observation (\ref{wrong}).

Now, it is clear how we could ameliorate the situation. Namely, to keep the Dirac string
unphysical we should remove the string field from the expression (\ref{delta}) which arises
automatically if we use the propagators (\ref{propagators}) following from the Zwanziger Lagrangian.
Thus, one should change ${\bf H}_1\cdot{\bf H}_2$ in the expression (\ref{delta}) to
${\bf H}_{1,rad} \cdot {\bf H}_{2,rad}$ so to say by hand. The justification is that
we should remove the effect of the string field from any observable. Therefore the result
(\ref{sch}) is inconsistent due to the interaction of virtual particles with Dirac string.
The correct propagator $\langle B_{\mu}B_{\nu}\rangle$ at the one-loop level is
\beq
\label{correct}
\langle B_{\mu}B_{\nu}\rangle(k) ~=~{\delta_{\mu\nu}\over k^2} (1+L)\,.
\eeq
In particular, there is no Dirac string self-energy renormalization.

One might wonder, how it happens that the couplings in the electric and magnetic potential
run in opposite ways. Indeed, now we reduced the product ${\bf H}_1\cdot{\bf H}_2$ to exactly
the same form as the product ${\bf E}_1\cdot {\bf E}_2$ in case of two electric charges (since
the radial magnetic and electric fields are the same, up to a change of the overall constants).
The resolution of the paradox is that the renormalizations of the electric and magnetic fields
are indeed similar in the language of the Lagrangian. However, the small
corrections to the Lagrangian and Hamiltonian are related as:
\beq\label{delta1}
\delta L~=~-\delta H.
\eeq
Since ${\bf E}^2$ and ${\bf H}^2$ enter with the same sign into the expression for the Hamiltonian
and with the opposite signs into the Lagrangian, Eq.~(\ref{delta1}) implies that the running of
the couplings in the electric $V_{e}$ and magnetic $V_m$ potentials is opposite in sign.
Which is, of course, in full agreement with expectations since $V_e\sim g^2$ and $V_m\sim g^{-2}$.

Return now to the evaluation of first radiative corrections to the potential (\ref{tree}).
For the sake of definiteness, we take
$n^a=\delta^{a,3}$ and consider the Feynman gauge. The consistency of this gauge with
Eq.~(\ref{F-on-Sigma}) can be established in the same way as for
the background gauge condition above (see the discussion after Eq.~(\ref{n-deviates})).
Since Eq.~(\ref{F-on-Sigma}) is satisfied, the dependence on the field $\chi^a$ can be
ignored\footnote{
Note that this approach might be too naive and dynamics of $\chi^a$ should
be considered more carefully. We leave this question for future investigation.
}. Then the free field Lagrangian corresponding to (\ref{dual-L}) is:
\beq
\label{free}
L^0 ~=~ \frac{1}{2}( m \cdot [\diff\wedge A] )^2 ~+~ \frac{1}{2}( m \cdot [\diff\wedge B] )^2
~-~ i \, ( m \cdot [\diff\wedge B] ) ( m \cdot \dual{[\diff\wedge A]} )~+~
\eeq
$$
~+~ \bar{W}_\mu (-\diff^2) W_\mu ~+~ \frac{1}{2}[ (\diff B)^2 + (\diff A)^2]
~+~ i \bar{c}\,\diff^2 c ~+~ i \bar{\zeta}\,\diff^2 \zeta ~+~ i \bar{\eta}\,\diff^2 \eta ~~+~~
i\,\frac{2\pi}{g}\,j\,B\,,
$$
where $A ~=~ A^3$, $W ~=~ 2^{-1/2}(A^1 + i A^2)$, $\bar{W} ~=~ 2^{-1/2}(A^1 - i A^2)$ and
$c$, $\bar{c}$, $\zeta$, $\bar{\zeta}$, $\eta$, $\bar{\eta}$  are  complex-valued ghost fields.
Furthermore, the interaction is given by:
\beq
\label{interaction}
L_{int}~=~ ig\,A_\mu\,\left[ \bar{W}_\mu (\diff W) - W_\mu (\diff \bar{W}) +
W_\nu \diff_\mu\bar{W}_\nu - \bar{W}_\nu \diff_\mu W_\nu \right] ~+
\eeq
$$
+~ 2ig\,[\diff\wedge B]\, \dual{\mathrm{Im}}(...) ~+~
g\, \mathrm{Im}(...)\,\left\{\;
[m\wedge (m\cdot [\;\diff\wedge A + i \dual{\diff\wedge B}\;]\,)\,] -
[\;\diff\wedge A + i \dual{\diff\wedge B}\;] \;\right\} ~+
$$
$$
~+~ g\,[ \bar{\zeta}(A\diff)\zeta - \bar{\eta}(A\diff)\eta ] ~+~
ig\,[ \bar{c}(W\diff)\zeta + \bar{c}(\bar{W}\diff)\eta - \bar{\eta}(W\diff) c -
\bar{\zeta}(\bar{W}\diff) c ]~+~ ...
$$
where $ \mathrm{Im}(...)_{\mu\nu} = \mathrm{Im}(W_\mu \bar{W}_\nu)$ and dots denote terms
quartic in the fields.

The first quantum correction to the potential (\ref{tree}) is due to the renormalization
of $\langle BB \rangle$ propagator. As is evident from the interaction Lagrangian (\ref{interaction})
in $O(g^2)$ order only $W\bar{W}$ vacuum polarization effects should be considered.
Therefore, we can use the Eq.~(\ref{Zw-constraint})
to argue that the second term in the middle line of Eq.~(\ref{interaction}) describes  in fact the
interaction of virtual $W\bar{W}$ pairs with Dirac string. In order to maintain the unphysical
nature of the string such an interaction should be excluded as we have already discussed in detail.
Thus we simply discard this term in the Lagrangian (\ref{interaction}).

\vspace{3mm}

\begin{figure}[t]
\centerline{\psfig{file=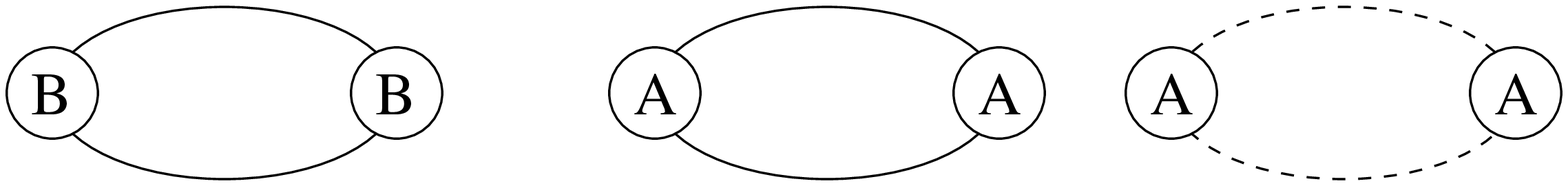,width=0.8\textwidth,silent=,clip=}}
\leftline{\hspace{0.2\textwidth} a) \rule{0mm}{6mm}\hspace{0.4\textwidth} b)}
\noindent\fontsize{11}{13.2}\selectfont
Figure 1: The diagrams relevant for the calculation of one-loop correction to
$\langle B_\mu B_\nu\rangle$ propagator. The solid and dashed lines represent
$W\bar{W}$ and ghost particles respectively.
\fontsize{12}{14.4}\selectfont
\end{figure}

Then the evaluation of the radiative correction to the $\langle B_\mu B_\nu \rangle$ propagator
reduces to the calculation of three diagrams shown on the Fig.~1, since vacuum loop with both
$B W\bar{W}$ and $A W\bar{W}$ vertices vanishes identically. In the leading log approximation
these diagrams give
\beq
\mbox{ Fig.~1a} ~=~ \delta^{\mu\nu}\; k^2\; \frac{g^2 L}{(4\pi)^2} \; (-8)\,,
\eeq
\beq
\label{AA}
\mbox{ Fig.~1b} ~=~ \delta^{\mu\nu}\; k^2\; \frac{g^2 L}{(4\pi)^2} \; (-\frac{2}{3})\,,
\eeq
where we have neglected all the terms proportional to $k_\mu$, $k_\nu$ and
$ L = \ln( \Lambda^2_{UV}/k^2)$.
The $\langle BB \rangle$ propagator with account for the quantum correction (\ref{AA}) is therefore
\beq
\label{AA-1}
\frac{-i}{k^2}\,\varepsilon_{\mu\mu'\sigma\tau}\,\frac{m^\sigma k^\tau}{(km)}\; \cdot \;
\delta^{\mu'\nu'}\; \frac{g^2 L}{(4\pi)^2}\,k^2\,(-\frac{2}{3})\; \cdot \;
\frac{i}{k^2}\,\varepsilon_{\nu'\nu\sigma'\tau'}\,\frac{m^{\sigma'} k^{\tau'}}{(km)}~=
\eeq
$$
=~ \frac{\delta_{\mu\nu}}{k^2} \cdot (1 ~-~ \frac{g^2 L}{(4\pi)^2}\,\frac{2}{3}) ~+~
\frac{g^2 L}{(4\pi)^2}\,\frac{2}{3}\cdot\frac{1}{(km)^2}(\delta_{\mu\nu} ~-~ m_\mu m_\nu)
$$
and is essentially of the form (\ref{sch}). As we have argued above the one-loop result
of the type (\ref{sch},\ref{AA-1}) is in fact wrong since it includes the interaction
of virtual particles with string singularity. The correct treatment of Dirac sheet leads
to Eq.~(\ref{correct}) instead of (\ref{sch}). Therefore,
the $\langle BB \rangle$ propagator at the one-loop level is
\beq
\langle B_\mu B_\nu \rangle ~=~
\frac{\delta_{\mu\nu}}{k^2}\cdot (1 - \frac{g^2 L}{(4\pi)^2}\;(8-\frac{2}{3})) ~=~
\frac{\delta_{\mu\nu}}{k^2}\cdot (1 - \frac{g^2 L}{(4\pi)^2}\;\frac{22}{3})\,.
\eeq
Thus we have reproduced the running of the coupling in the inter-monopole potential (\ref{tree}).

The general conclusion which can be drawn from consideration of the radiative corrections
is that the Lagrangian approach produces reasonable physical results only if it is supplemented
by particular rules of handling the singular magnetic field of the Dirac strings attached
to the monopoles. The rule is that we should always subtract the contribution
of the strings, even if it is not vanishing within the perturbation theory.
Essentially, the rule is that the product ${\bf H_1\cdot H}_2$
should not include the piece ${\bf H}_{rad}\cdot {\bf H}_{string}$.

To justify this rule we have to go actually
beyond the Zwanziger Lagrangian and consider a regularization procedure which
takes into account the Dirac strings.
It is worth emphasizing that within the lattice regularization
the term ${\bf H}_{rad}\cdot {\bf H}_{string}$ is indeed absent.
Let us recall the reader that on the lattice the Dirac string corresponding to $|Q_m|=1$ monopole
pierces negative plaquettes. This is true in the limit $g^2\to 0$, or $a\to 0$.
If one looks for small deviation from $U_p=-1$, then the action does not contain terms,
which are linear in perturbations. Note that this will be not true
for the expansion around arbitrary $U_p \ne \pm 1$. This result means
in turn that in the continuum limit there is no term ${\bf H}_{rad}\cdot {\bf H}_{string}$.
Thus, our naive removal of the effects of the Dirac string from the radiative corrections,
is justified by the lattice regularization.

So far we considered only graphs with two insertion of the external monopole field.
At this level, the value of the magnetic charge is not important.
For example, the same arguments would go through without
change if we started with, say, monopoles with $Q_m=2$.
But such monopoles are unstable \cite{Brandt-Neri} and this is a much
more drastic effect than the would-be running of the coupling.
The instability is revealed by looking for the negative modes in the external
field, i.e. considering infinitely many insertions
of the external field. In particular,
the instability of $|Q_m|=2$ monopole is due to the strong interaction
of the spin of charged gluons with the magnetic field of the monopole.
As a result of such interaction, the gluons fall onto the center of $|Q_m|=2$ monopole
thus invalidating any classical consideration.
For the monopoles with $|Q_m|=1$ there is no instability.
Nevertheless, one could question whether the system of a monopole and antimonopole
is also stable. There are no analytical ways to study the stability in this case.
Numerical simulations \cite{main} show, however, no sign of instability, as one would expect.

\section{$|Q_m|=2$ Monopoles and Effective Theories.}
\label{Effective-Model}

So far we discussed the fundamental monopoles $|Q_m|=1$ which can be visualized as classical infinitely
heavy objects. Because of the infinite mass,
these monopoles can be used only as probes of the QCD vacuum but play
no dynamical role by themselves. The $|Q_m|=2$ monopoles at small distances, as is mentioned above,
are essentially trivial, since in the ultraviolet
they are indistinguishable from singular gauge transformations.
On the other hand, there exist very simple
arguments that they can play dynamical role at distances $r \sim \Lambda_{QCD}^{-1}$.
Indeed, as far as the fundamental Lagrangian is concerned,
the only role of the quantum corrections is the
running of the non-Abelian coupling. In particular, if we consider a lattice coarse enough then
$g_{SU(2)}$ becomes of order unity. Obviously, the same coupling governs the physics associated with any
$U(1)$ subgroup of the $SU(2)$. However, if the coupling $g_{U(1)}$ becomes of order unity,
then there is a phase
transition associated with the monopole condensation \cite{polyakov}.
Thus, it is very natural to assume that the monopole condensation occurs also in QCD since the running of
the coupling allows to scan the physics at all the values of $g_{SU(2)}$ until one runs into a phase
transition.

The $|Q_m|=2$ monopoles condense in the vacuum and their physics is commonly described
by field theories incorporating the Higgs mechanism \cite{baker,brambilla,ichie,kondo}.
This kind of models, however, fail to reproduce the Casimir scaling \cite{greensite}
and in this section we confront with this test our model
with $U(1)$ dual gluon matching color gluons in the adjoint representation

In order to have a more quantitative description of the problem let us consider the partition function
(\ref{con-PF}) which by assumption correctly describes the $SU(2)$ lattice gauge theory in the
continuum limit.
What is of particular importance is that in this representation there  is no much difference between (\ref{con-PF})
and the expectation value (\ref{dual-H}), since for every configuration of $\Sigma$,  Eq.~(\ref{con-PF})
essentially coincides with (\ref{dual-H}). Since it is known
\cite{main} that (\ref{con-PF}) depends only on the boundaries of $\Sigma$, we can directly apply
the results of the previous section and represent (\ref{con-PF}) in the form:

\beq
\label{Z-1}
Z~=~\int \cD j \; \cD A \cD B \cD \chi \! \int\limits_{n^2=1} \!\!\! \cD n \;
\exp\{\; -\int L(A,B,\chi,n) ~+~ i \frac{4\pi}{g} \int j\,B  \;\}\,,
\eeq
where $L(A,B,\chi,n)$ is given by (\ref{dual-L})
and $\int \cD j$ denotes a well known measure of integration over closed
world-lines of $|Q_m|=2$ monopoles.

The next step would be to perform the summation over monopole currents $j$.
The integral $\int \cD j$ may exactly be rewritten as a functional integral over
complex scalar field $\Phi$ \cite{samuel}:
\beq
\int \cD j \exp\{\;- S(j) +  i \int j H \;\} ~=~
\int \cD \Phi \exp\{\; -\int [ \frac{1}{2} |(\diff ~+~ i H )\Phi|^2 + V(|\Phi|)\,]\;\}\,,
\eeq
where $H_\mu$ is an arbitrary real-valued vector field. The concrete form of the potential $V(|\Phi|)$
depends crucially on the current-current interaction $S(j)$. In particular, for $S(j)=0$ the potential
vanishes, while more sophisticated current interactions lead to non-trivial $V(|\Phi|)$.
For instance, the simplest form $S(j) \sim j^2$ results in the $|\Phi|^4$ term in the effective Lagrangian
for the field $\Phi$.

Unfortunately, the exact form of the current-current interaction and hence the potential $V(|\Phi|)$
is so far unknown. Nevertheless, we may assume without loss of generality that in the
infrared region the potential $V(|\Phi|)$ is of the Higgs type:
\beq
V(|\Phi|) ~=~ \lambda \left( |\Phi|^2 - \eta^2 \right)^2\,,
\eeq
where $\lambda$ and $\eta$ are phenomenological constants. Of course, the vacuum expectation value
of the Higgs, or monopole field is of order $\Lambda_{QCD}$. Therefore, we get the following effective
theory of $SU(2)$ gluodynamics
\beq
\label{Z-eff}
Z~=~\int \cD A \cD B \cD \Phi \;  \cD \chi \! \int\limits_{n^2=1} \!\!\! \cD n \;
\exp\{\; -\int \mathrm{d}^4x L_{\mathrm{eff}} \;\}\,,
\eeq
\bea{cc}
\label{L-eff}
\displaystyle{ L_{\mathrm{eff}} ~=~ } &
	\displaystyle{\frac{1}{4}( F^a_{\mu\nu})^2 ~+~
	\frac{1}{2} \Bigr[ 
		\Bigr(
			m\cdot [ \diff\wedge B  ~-~
			i\;\dual{F^a n^a} ~-~
			\varepsilon^{abc} \chi^a n^b \dual{F}^c \, ]
		\Bigr)_\mu\,
	\Bigr]^2 ~+} \\
\rule{0mm}{7mm}                      &
	\displaystyle{ +~ \frac{1}{2} |(\diff + i\frac{4\pi}{g} B)\Phi|^2 + V(|\Phi|)}\,,
\eea
which is presumably valid at low energies.

Despite its apparent simplicity, Eq.~(\ref{L-eff}) is highly speculative. Namely, it unifies
point-like gluons described by the fields  $A^a$, $B$ and $\Phi$ which
is an effective scalar field. One may justify the use of (\ref{L-eff}) by assuming that the effective
size of the monopoles with $|Q_m|=2$ is in fact numerically small,
although generically it is of order $\Lambda_{QCD}$ (see also \cite{shuryak}).

What is also specific about the Lagrangian (\ref{L-eff}) is that the dual gluon is a $U(1)$ gauge boson.
The color symmetry is maintained by averaging over all possible embeddings of the (dual)
$U(1)$ into $SU(2)$, see the discussion in section~\ref{Z_2-1}. The confinement mechanism inherent
to (\ref{L-eff}) is the formation of the Abrikosov-Nielsen-Olesen string \cite{abrikosov}
which can be considered already on the classical level and which we discuss in detail below.
More generally, the Lagrangian
(\ref{L-eff}) exhibits the {\it Abelian dominance} in the confining region which is the dominance of
Abelian-like field configurations in the full non-Abelian theory. This dominance is common
to all the realizations of the dual-superconductor model of confinement \cite{confinement}
and is strongly supported by the lattice data \cite{map}.  What we avoid, however, is the
{\it breaking} of $SU(2)$ to $U(1)$ which is inherent to the models \cite{baker} which start with
the dual gluons in the adjoint representation and then add effective isospin one Higgs fields.

{}From the theoretical point of view, the most important limitation in
the use of (\ref{L-eff}) is that it can be consistently treated on the classical level only.
The difficulty to extend it to the quantum level is due to the Dirac strings.
Indeed, perturbatively the Dirac veto is violated for virtual particles (see Section \ref{Z_2-2}).
When the monopoles condense the Dirac strings are filling the whole of the vacuum
and there are no known ways to rectify the perturbation theory. In case of the Abelian Higgs model,
it is even more convenient to use the dual language when the Dirac strings are attached to the
electric charges. Then, to respect the Dirac veto one should impose the condition
that the Higgs field vanishes along a line connecting charges \cite{ss}.
The static Abrikosov-Nielsen-Olesen string satisfies this constraint \cite{abrikosov}.
However, there is no known complete set of solutions satisfying this boundary condition.

To reiterate the point,
in the standard field theory we can reproduce, for example, interaction of two static
charges in two different but equivalent ways. Classically, we solve the
equations of motions for the sources which are conserved currents. Quantum mechanically,
we study the correlator $\langle A_\mu(x) \, A_\nu(y)\rangle$
and substitute the vacuum zero-point
fluctuations for the potentials $A_\mu(x)$. In case of the Abelian Higgs model
with both electric and magnetic charges the quantum propagator cannot be constructed
explicitly. Moreover, there is no doubts that it would involve the physics of strings.
Thus, we are confined to the former approach when only classical static solutions
are known.

How far can we go within the classical approximation? Strictly speaking,
the classical approximation by itself does not allow to make any physically
interesting predictions. The point is that we are considering now the interaction
of quarks which are degenerate in color. In particular, we are interested in the
potential energy for heavy quarks in the color singlet state.
This notion cannot be formulated entirely in terms of the classical field theory.
Quantum mechanically, the projection to the color singlet state
is implemented by considering the Wilson loop.
It would be, therefore, desirable to formulate classical solutions in terms of
propagators.

Consider first the $U(1)$ Zwanziger Lagrangian (\ref{Zw-action}) where the magnetic
current is constructed on a scalar field which possesses also the Higgs-like
self-interaction. Formally the $\langle A_\mu\, A_\nu \rangle$ propagator can be easily found
(see, e.g., \cite{ss} and references therein):
\beq\label{bl}
\langle A_{\mu}\, A_{\nu}\rangle~=~
{1\over k^2+m_V^2}\big(\delta_{\mu\nu}-{m_V^2\over (k m)^2}
(\delta_{\mu\nu}-m_{\mu}m_{\nu}) + \, ...\, \big)\,,
\eeq
where dots denote the longitudinal terms.
Taken at face value, the propagator (\ref{bl}) makes, however, no sense since it has
a double pole in the $(k m)$ variable and thus depends explicitly on the
direction of Dirac string. All these pathological features manifest
violation of the Dirac veto in the formal derivation of
the propagator (\ref{bl}). The right way is to solve the classical equations
anew supplementing them by the boundary condition of vanishing of the Higgs field along a
line connecting the external charges \cite{ss}. However, one can introduce instead formal
rules of integrating the double pole out and derive the potential energy in coordinate space
corresponding to an exchange of a single $A$-field quantum \cite{suzuki}. In the London limit,
$m_H \gg m_V$ the result coincides with the solution of the classical equations of motion.
The tension of the string, which connects the external charges $\pm N g$ is:
\beq
\sigma_{Ab}( N g )~\approx~ N^2 \, {\pi\over 2 g^2} \, m_V^2 \, \ln{m_H^2\over m_V^2}\,.
\eeq
For an arbitrary
relation between $m_H$ and $m_V$ one should allow for an adjusting overall factor
\cite{ss}.

Although such a use of the propagator may look awkward\footnote{
Alternatively, we could use the language of quantum mechanics. Indeed, our problem is similar to,
say, the problem of hyperfine splitting. The role of the Wilson loop is then to ensure that the
external $q\bar{q}$ pair is in the $T=0$ state. While the knowledge of the classical solution
allows to evaluate the potential in case when the sources have a definite (and opposite) $T_3$.
Averaging of the potential over the wave function gives the final result.
}, it teaches us once again
that the classical solution corresponds to a single quantum exchange. Note that
the double exchange would result in infrared divergent expression. However, the rule is
that one should not iterate the quantum exchanges to reproduce a classical solution
(for a similar analysis of the instanton solution see \cite{zakharov}).

Next, consider classical solutions
corresponding to the Lagrangian (\ref{L-eff}). Clearly, it has the same, Abelian-type
solutions for any choice of the vector $n^a$. We can therefore supplement the propagator (\ref{bl})
by the Kroneker symbol $\delta^{a,b}$ in the color space and reduce the effect
of the classical solutions to a single quantum exchange. Or, we can fix
$n^a = \delta^{a,3}$ and consider the Wilson loop in the single-quantum exchange approximation.
Note that all these remarks apply also to a single-gluon exchange as well.
In this sense, there is no much specific about the Abrikosov-Nielsen-Olesen
string exchange. Thus the criticism of Ref.~\cite{greensite} does not apply in our case. We do have
a $U(1)$ dual gluon, but no $U(1)$ subgroup is singled out for the ordinary gluons.
The fixation of $U(1)$ is a matter of gauge fixing and convenience.

What is specific for the one-gluon exchange is that the overall coefficient
is fixed by the condition that the gluon is coupled via the generators of the gauge group.
In case of the string, the string tension which plays the role of the coupling
is a dynamical quantity depending on the parameters of the theory,
that is $m_{H,V}$.
In this way we derive linear potential for heavy quarks in any representation $T$.
As for the coefficient in front of the linear rising term, it should be now calculated explicitly
as function of $m_V,m_H$:
\beq
\sigma_T~=~{1\over 2T+1}\,\sum\limits_{T_3 = -T}^{T} \sigma_{Ab}( 2 T_3)\,,
\eeq
where $\sigma_{Ab}$ denotes the string tension, calculated in  pure Abelian Higgs model with
the same values of $m_{H,V}$ and for external charges of magnitude $\pm 2 T_3$.
In the London limit, $m_H\gg m_V$, the string tension $\sigma_{Ab}(2 T_3) \sim T_3^2$ and we
reproduce the Casimir scaling since
$$
{1\over 2T+1}\,\sum\limits_{T_3 = -T}^{T} T_3^2 ~=~ \frac{1}{3} \, T(T+1) \,.
$$

So far we discussed only embedding of the $U(1)$ into $SU(2)$. As far as the physics of large
distances is concerned, there is a subtle point how to define the $U(1)$ dynamically. As is
mentioned above the $U(1)$ arises in lattice simulations through gauge fixing \cite{u1}.
The monopole properties depend on the particular choice of $U(1)$. The Maximal Abelian projection
turns most successful as far as the interaction of the heavy quarks in the fundamental representation
is concerned \cite{map}.

It is plausible to assume that the Lagrangian (\ref{L-eff}) with $n^a = \delta^{a,3}$
describes the data in this gauge,
although at present there are no theoretical arguments in favor of this. Moreover, if this assumption
is valid, then a new phenomenological problem arises. Namely, the structure of the confining string
for the quarks in the fundamental representation is best described in the
Bogomolny limit \cite{bogomolny}, $m_H\approx m_V$. While the Casimir scaling can be reproduced only
when $m_H\gg m_V$. Thus, an overall fit can be obtained only at price
of compromising the quality of the fits. Which might be not surprising since we are
relying on the classical approximation to fit the data obtained via the lattice simulations
which are fully quantum.

\section{Conclusions.}
\label{Conc}

The main objective of this paper was to develop an
approach to gluodynamics which would allow to describe interaction both
with color and magnetic charges. We considered first external magnetic charges
which can be introduced via the 't~Hooft loop. We used the Zwanziger-type approach
which formally starts with two different fields for ordinary and dual gluons,
interacting with the color and magnetic charges, respectively.
The doubling of the degrees of freedom is avoided then by imposing constraints.
Our central point is that it is natural to combine ordinary gluons in the adjoint representation
with $U(1)$ dual gluon. The reason is that the external monopoles are in fact Abelian objects
and do not constitute representations of the non-Abelian group.
The fixation of the $U(1)$ is a kind of new gauge freedom. In particular it can be fixed
to be space-time independent. The overall non-Abelian symmetry is maintained either by
averaging over all embeddings or by confining the consideration only to gauge-invariant
quantities, i.e., the Wilson loops.

We checked the viability of the approach by considering the running of the coupling
governing the interaction of two external monopoles at short distances.
We have shown that the coupling runs as expected, i.e., the Dirac quantization
condition is preserved with account of the radiative corrections.
To derive the running of the coupling it is essential to respect for the virtual particles
the Dirac veto which forbids any direct interaction with the Dirac strings attached
to the external monopoles. In the field theoretical language the veto is manifested
through a particular ultraviolet regularization. This regularization can either
be postulated or derived by analyzing the continuum analog of the 't~Hooft loop.

We turned then to consideration of large distance physics which is commonly described
in terms of monopole condensation. Here, the monopoles with a double charge
are meant which are not point-like, unlike the external monopoles introduced
via the 't~Hooft loop. What unifies the description of the both types of the monopoles
is that the dual gluon is a $U(1)$ boson in both cases. This does not imply
any violation of the color $SU(2)$ either at short or large distances.
Classical, Abrikosov-type solutions to the effective theory produce linear potential
for heavy quarks in any representation of the color group and allow, in particular,
to incorporate the Casimir scaling. The use of the effective theory is limited, however,
by the necessity to impose the Dirac veto for virtual particles. In practice, there
is no known way to go beyond the classical approximation. At the quantum level,
the physics of strings seems to be indispensable.

\section*{Acknowledgments}

M.N.Ch. and M.I.P. acknowledge the kind hospitality of the staff of the
Max-Planck Institut f\"ur Physik (M\"unchen), where the work was initiated.
Work of M.N.C., F.V.G. and M.I.P. was partially supported by grants RFBR
99-01230a, INTAS 96-370, Monbushu grant and CRDF award RP1-2103.


\end{document}